\documentclass[11pt]{article}

\AtBeginDocument{%
	\providecommand\BibTeX{{%
			\normalfont B\kern-0.1em{\scshape i\kern-0.25em b}\kern-0.1em\TeX}}}

\usepackage{subcaption}

\usepackage{amsmath,amssymb,amsfonts}
\usepackage{graphicx}
\usepackage{textcomp}
\usepackage{xcolor,tabularx}
\usepackage{color, colortbl}
\usepackage{pifont}
\usepackage{mathtools}
\usepackage[margin=1in]{geometry}
\usepackage{amsmath,amssymb,amsthm}
\usepackage{booktabs}
\usepackage{graphicx,color}
\usepackage[T1]{fontenc}
\usepackage{tikz}
\usetikzlibrary{arrows,positioning,automata,calc,shapes,intersections}
\usepackage[linesnumbered,ruled,vlined]{algorithm2e}
\usepackage{url}

\usepackage[most]{tcolorbox}
\usepackage{empheq}
\newcommand\myeq{\stackrel{\mathclap{\normalfont\mbox{def}}}{=}}

\newtcbox{\mymath}[1][]{%
    nobeforeafter, math upper, tcbox raise base,
    enhanced, colframe=blue!30!black,
    colback=blue!30, boxrule=1pt,
    #1}

\newcommand{\ab}{{\sc AntiBenford}\xspace}

\newcommand{\spara}[1]{\smallskip{\bf #1}}

\newcommand{\mycomment}[1]{}

\newcommand{\hide}[1]{}

\newcommand{\field}[1]{\mathbb{#1}} 

\makeatletter
\newcommand*\bigcdot{\mathpalette\bigcdot@{.5}}
\newcommand*\bigcdot@[2]{\mathbin{\vcenter{\hbox{\scalebox{#2}{$\m@th#1\bullet$}}}}}
\makeatother

\theoremstyle{plain}
\newtheorem{theorem}     {Theorem}

\newtheorem{proposition} {Proposition}
\newtheorem{corollary}   {Corollary}
\newtheorem{definition}  {Definition}
\newtheorem{problem}     {Problem}

\newtheorem*{problem*}     {Problem}

\newcommand{\Prob}[1]{{{\bf{Pr}}\left[{#1}\right]}}
\newcommand{\Mean}[1]{{\mathbb E}\left[{#1}\right]}
\newcommand{\Var}[1]{{\mathbb Var}\left[{#1}\right]}

\newcommand{\squishlist}{
 \begin{list}{$\bullet$}
  {  \setlength{\itemsep}{0pt}
     \setlength{\parsep}{3pt}
     \setlength{\topsep}{3pt}
     \setlength{\partopsep}{0pt}
     \setlength{\leftmargin}{2em}
     \setlength{\labelwidth}{1.5em}
     \setlength{\labelsep}{0.5em}
} }
\newcommand{\squishlisttight}{
 \begin{list}{$\bullet$}
  { \setlength{\itemsep}{0pt}
    \setlength{\parsep}{0pt}
    \setlength{\topsep}{0pt}
    \setlength{\partopsep}{0pt}
    \setlength{\leftmargin}{2em}
    \setlength{\labelwidth}{1.5em}
    \setlength{\labelsep}{0.5em}
} }

\newcommand{\squishdesc}{
 \begin{list}{}
  {  \setlength{\itemsep}{0pt}
     \setlength{\parsep}{3pt}
     \setlength{\topsep}{3pt}
     \setlength{\partopsep}{0pt}
     \setlength{\leftmargin}{1em}
     \setlength{\labelwidth}{1.5em}
     \setlength{\labelsep}{0.5em}
} }

\newcommand{\squishend}{
  \end{list}
}

\begin{document}
	\pagenumbering{gobble}
	
	\title{AntiBenford Subgraphs:  \\ Unsupervised   Anomaly Detection in Financial Networks}

	\author{
 Tianyi Chen \\ Boston University \\ ctony@bu.edu 
 \and 
  Charalampos E. Tsourakakis \\ 
  Boston University \\ ctsourak@bu.edu
    }

	\maketitle

	\begin{abstract}
		Benford's law describes the distribution of the first digit of numbers appearing in a wide variety of numerical data, including tax records, and election outcomes, and has been used to raise "red flags" about potential anomalies in the data such as tax evasion. In this work,  we ask the following novel question:

\begin{quotation}
	Given a large transaction or financial graph, how do we find a set of nodes that perform many transactions among each other that  deviate significantly from Benford's law?  
\end{quotation}

We propose the \ab subgraph framework that is founded on well-established statistical principles. Furthermore, we design an efficient algorithm that finds \ab subgraphs in near-linear time on real data. We evaluate our framework on both real and synthetic data against a variety of competitors. We show empirically that our proposed framework enables the detection of anomalous subgraphs in cryptocurrency transaction networks that go undetected by state-of-the-art graph-based anomaly detection methods. Our empirical findings show that our \ab framework   is able to mine anomalous subgraphs, and provide novel insights into financial transaction data.  The code and the datasets are available at \url{https://github.com/tsourakakis-lab/antibenford-subgraphs}.

	\end{abstract}

\section{Introduction}
\label{sec:introduction}

\begin{figure*}
	\centering
	\begin{tabular}{cc}
		\includegraphics[width=0.35\textwidth]{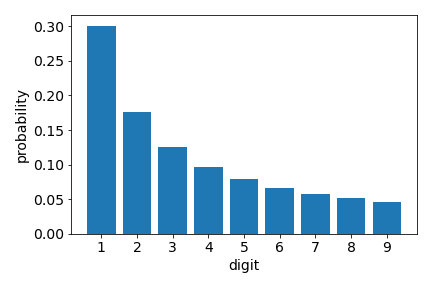} &	\includegraphics[width=0.55\linewidth]{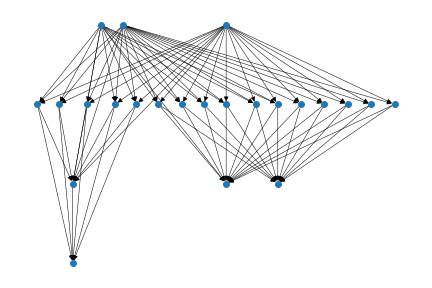} \\
		(a) &	(b) 

	\end{tabular} 
	\caption{\label{fig:enron_intro} (a) Benford's First Digit Law.  (b) AntiBenford subgraph  from the Ethereum financial network during January 2018.  The transaction amounts starting with digit 5 have high frequency. The resulting subgraph resembles a tripartite-like clique where the amounts end up in a few accounts. For more, see Section~\ref{sec:exp}.  }
\end{figure*}

Anomaly detection is a broad term describing methods that detect patterns that deviate from normal behavior \cite{chandola2009anomaly}. The development of an anomaly detection method is application-dependent, and requires an understanding of what constitutes ``normal''. For example, consider the following question of interest to revenue services:  how do we detect tax evasion and fraud?  What information can we use in order to raise ``red flags'' about certain accounts that will undergo additional control? Benford's law, aka Newcomb-Benford law \cite{benford1938law,newcomb1881note}, states that the first digit of numbers  appearing in natural datasets, such as addresses, areas, stock quotations, town populations, is not uniformly distributed. Specifically, the fraction of numbers starting with digit $d$ is not the uniform $\frac{1}{9}$ ($d=1,\ldots,9$),    but instead follows the monotone decreasing function of $d$, i.e., $\log_{10}(1+1/d)$. This distribution is shown in Figure~\ref{fig:enron_intro}(a).   Benford's law has been used numerous times successfully in practice to raise red flags that eventually detect fraud, see for instance \cite{roukema2014first,drake2000computer,Sambridge2010, Luis2011, Cerioli106,vicic2021benford}.  

In this work we are interested in developing novel unsupervised methods for detecting anomalous subgraphs in financial and transaction networks. Despite the large toolbox of unsupervised anomaly detection methods~\cite{goldstein2016comparative,tsai2007unsupervised}, the task of designing novel, well-performing unsupervised methods remains a great challenge.  Furthermore, the case of transaction networks is special for (at least) two major reasons. First, transaction networks are knowledge graphs, with a lot of important information available for the nodes and edges; e.g., we know that if a node has a PO box only, or resides in what is called {\it high risk} country,  the conditional probability of being part in an illicit scheme is higher than some node that has a standard bank account in a low risk country.  These special characteristics can play a key role in detecting anomalies, but have not been taken account largely by the research community. One of the key reasons behind this fact is the scarcity of real-world banking data, as well as the natural hesitation of financial institutions to publish more on their anomaly detection systems, despite the existence of dedicated venues (e.g., Journal of Money Laundering Control). Despite the scarcity of publicly available bank transaction datasets due to privacy reasons, the wide-spread use of cryptocurrencies has enabled the collection of interesting financial datasets. For instance, in a Bitcoin transaction network, nodes correspond to transactions, and edges to Bitcoins flow between transactions \cite{weber2019anti}.  

Finding subgraphs that correspond to illicit behavior such as malware, money laundering, ransomware, and Ponzi schemes is a challenging task. Towards this goal, we propose a novel graph based anomaly detection method that leverages two components: (i) Benford's law, and (ii) dense subgraph discovery, a major area of graph mining. The latter has been used widely in a variety of tasks ranging from finance to biology, and also anomaly detection, e.g., ~\cite{eswaran2018spotlight,liu2017holoscope,li2020flowscope,hooi2016fraudar,mitzenmacher2015scalable}. The latter is due to the intuitive fact that in certain applications statistically unlikely dense clusters correlate well with anomalous, and possibly malicious behavior.   .  

In this work we propose the \ab subgraph anomaly detection problem, stated informally as follows: 

\begin{tcolorbox}
	\begin{problem}
		\label{prob1} 
		\normalfont
		\spara{(Informal) Anti-Benford subgraphs.} Given a weighted network $G(V,A,w), w:A \rightarrow \field{R}^+$ where each arc $e=(u,v) \in A$ corresponds to a transaction, and $w(u,v)$ is the amount of money $u$ sends to $v$, find a subset of nodes $S\subseteq V$ such that (i) it induces on average many edges, and (ii) whose weights' first digit	empirical histogram deviates from Benford's law significantly, in a statistical sense. 
	\end{problem}
\end{tcolorbox}

\noindent Intuitively, Antibenford subgraphs are dense subgraphs whose edges have an empirical distribution significantly different from Benford's law. The edge density constraint (i) biases the output towards a set of nodes that in some way are likely to know each other,  or in the context of cryptocurrency financial networks, they may even correspond to wallets held by a single person or a small group of people who coordinate their actions. The second constraint (ii) biases the output to include edges whose weights ``violate'' Benford's law in the following sense; if the edges' first digits are independent and identically distributed (iid) random variables (RVs)  according to Benford's law, then  their distribution is a multinomial. There exists a variety of  {\it goodness-of-fit} tests for testing such hypotheses, e.g.,  \cite{hoeffding1965asymptotically,smith1981approximating}. As we show experimentally in Section~\ref{sec:exp}, combining  constraints (i), and (ii) results in finding interesting, and  statistically significant anomalous subgraphs. It is worth mentioning that we  empirically verify that  Ethereum blockhain transactions follow closely Benford's law, see Section~\ref{subsec:eth} for details.

Our contributions in this work include the following. 

\vspace{2mm}

$\bullet$ {\bf AntiBenford subgraphs.} We propose a novel unsupervised framework for detecting anomalies in financial networks. The generic task of developing such a framework  is known to be a challenging problem of great importance.  

\vspace{1mm}

$\bullet$ {\bf Mathematical and algorithmic analysis.} We propose a specific formulation for problem~\ref{prob1}, and design an  algorithm  that is algorithmically founded on dense subgraph discovery. Furthermore, we provide mathematical insights into the behavior of our objective using a probabilistic analysis.

\vspace{1mm}

$\bullet$ {\bf Experimental validation.}  Our method is able to uncover anomalous subgraphs in various settings.  For example, Figure~\ref{fig:enron_intro}(b) shows an AntiBenford subgraph found in the Ethereum network spanning a period of a month (Jan. 2018). The subgraph has an apparent tripartite structure, where the money flows from one layer to the next. Furthermore, the distribution of the first digit among this set of transactions deviates significantly from Benford's law, see Figure~\ref{fig:eth2018_19_dist}. The structure resembles a smurf-like\footnote{In banking, smurfing refers to splitting of a large financial transaction into multiple smaller transactions.} structure with respect to the layering, a subgraph that is known to appear in money laundering schemes \cite{lee2020autoaudit,starnini2021smurf}.    These findings are representative of what our proposed method can achieve in complex, networked data. Despite the remarkable success of prior graph-based anomaly detection methods, our findings verify that our framework provides information that is otherwise inaccessible.

\vspace{3mm}

\spara{Reproducibility.} Our code  is publicly available at \url{https://github.com/tsourakakis-lab/antibenford-subgraphs}. The paper is organized in the usual way, i.e., related work (Section~\ref{sec:related}), proposed method (Section~\ref{sec:proposed}), experiments (Section~\ref{sec:exp}), and conclusions (Section~\ref{sec:conclusion}).

	\section{Related Work}
\label{sec:related}

\spara{Graph Based Anomaly Detection} is an intensively active area of graph mining \cite{noble2003graph}, that attracts  interest from a diversity of  industrial and scientific applications. Such applications include security applications \cite{bhatia2020midas,chen2017practical}, anti-money laundering \cite{li2020flowscope}, spammers \cite{akoglu2010oddball,hooi2016fraudar,eswaran2018spotlight}, and fraud detection \cite{eberle2007discovering}.  We indicatively report some works in this area; the interested reader should confer the extensive survey on graph-based anomaly detection~\cite{akoglu2015graph}. $k$-cores have also been frequently used to detect anomalies in large-scale networks \cite{giatsidis2014corecluster,shin2016corescope}. Many frameworks tend to use both graph topology and node/edge attributes to identify anomalies. Noble and Cook~\cite{noble2003graph}  studied anomaly detection on graphs with categorical features by searching for graph sub-structures that occur infrequently.  The evaluation is based on the Minimum Description Length Principle (MDL). Davis et al.\cite{davis2011numeric} suggested an extension that is able to treat  both structural data and numeric attributes. OddBall is a scalable anomaly detection tool on weighted graphs~\cite{akoglu2010oddball}. To detect transaction frauds, Semenets et al. compress  each transaction as a star graph instance, and train  classifiers with vector representations generated based on frequent subgraphs~\cite{HerreraSemenets2015AFF}. Braun et al. constructed  a shop-card bipartite graph, and defined a suspiciousness score focusing on the bicliques \cite{Braun2017card}. Kumar et al. proposed REV2 to identify fraudulent users from rating platforms, by computing anomalous scores on nodes and edges with network and behavior properties~\cite{kumar2018rev2}. Agarwal et al. calculated $\chi ^2$ statistics to measure the statistical significance of subgraphs matching on uncertain graphs~\cite{Agarwal2020chisel}.   Recently, Nilforoshan and Shah
proposed an axiomatic-based method to find fraudsters on Snapchat's ad platform by incorporating characteristics that such fraudsters typically possess~\cite{nilforoshan2019slicendice}.  Mitzenmacher et al. \cite{mitzenmacher2015scalable} use the $k$-clique densest subgraph problem \cite{tsourakakis2015k} to find a citation ring in the patents citation network.  Graph anomaly detection has been applied to financial networks for applications including anti-money laundering, and fraud detection. Pham et al. applied unsupervised machine learning techniques including k-means clustering and SVM to detect anomalies on Bitcoin transaction network~\cite{Pham2016anomaly}. Recently Weber et al. published the  \textit{Elliptic dataset}, so far the largest labelled Bitcoin transaction dataset~\cite{weber2019anti}. Graph neural networks are used on financial networks, typically in a semi-supervised setting~\cite{Wang2019semi, weber2019anti, Zhang2021eFraud}.

Several dense subgraph discovery based methods have been proposed in the literature.  Fraudar is an algorithm for catching fraudulent sets of nodes in graphs that are dense and  that may attempt to  camouflage themselves by creating random connections with non-malicious/normal nodes \cite{hooi2016fraudar}.   HoloScope~\cite{liu2017holoscope}  is a  well-suited framework   for  bipartite graphs, that detects fraudulent attacks based on the graph topology, and also other information including temporal event bursts, drops, and rating scores. Their method relies on using singular vectors to detect well-connected communities that are then used as seeds to detect dense subgraphs based on node anomaly scores. FlowScope~\cite{li2020flowscope} aims to detect money laundering on transaction networks by detecting dense multi-partite structures.  Starnini et al. show that detecting smurfs using standard SQL joins with a time filter results in highly suspicious subgraphs of money laundering~\cite{starnini2021smurf}. Finally, EigenSpokes~\cite{Prakash2010eigen} uses the correlations between eigenvectors to detect near-bipartite cliques and near-cliques, see also \cite{Jiang2014infer} for further applications.

\spara{Dense subgraph discovery} is a major topic in graph mining \cite{gionis2015dense}.  Detecting dense subgraphs has a wide range of applications, ranging from anomaly detection \cite{li2020flowscope,hooi2016fraudar,eswaran2018spotlight,jiang2015general}, to  bioinformatics \cite{saha2010dense,fratkin2006motifcut}. A wide variety of mathematical formulations has been proposed in the literature. The archetypical dense subgraph is a clique. However, the maximum clique problem is not only NP-hard, but also strongly inapproximable, see \cite{hastad}.  Many other formulations that relax the clique requirement, and instead require that a large fraction of  all pairs of nodes are connected are also NP-hard, and typically hard to approximation. For instance, finding  an optimal quasi-clique \cite{kawase2018densest,tsourakakis2013denser} is also NP-hard  \cite{tsourakakis2015streaming}, but scalable solutions that work well on real-world datasets have been developed in recent years \cite{mitzenmacher2015scalable,konar2020mining}.



In sheer contrast to the quasi-clique formulations  with respect to the computational complexity,  stands the densest subgraph problem (DSP) that is solvable in polynomial time on graphs with non-negative weights \cite{gallo1989fast,khuller2009finding,goldberg1984finding}.  Given a graph $G(V,E,w), w:E \rightarrow \field{R}^+$  the {\em densest subgraph problem} (DSP) aims to find a  subset of nodes $S \subseteq V$ that maximizes the degree density $\rho(S)=\frac{w(S)}{|S|}$. Here, $w(S)$ denotes the total edge weight induced by $S$, i.e., $w(S) = \sum_{(i,j) \in E(S)} w(i,j)$. When the graph is unweighted, then the numerator $w(S)$ simply equals the number of induced edges $e(S)=|E(S)|$, and the degree density is simply half the average degree of the  subgraph $G(S)$ induced by $S$. When there exist negative weights, the DSP becomes NP-hard \cite{tsourakakis2019novel}.  We emphasize that the degree density objective is better suited for undirected, non-bipartite graphs; we  focus on this objective in this paper. More suitable objectives for other types of graphs have also been proposed in the literature, see  \cite{kannan1999analyzing,charikar2000greedy,khuller2009finding,ma2020efficient}.

The  DSP can be solved exactly using (parametric) maximum flows \cite{gallo1989fast,goldberg1984finding}. Since maximum flow computations are computationally expensive \cite{orlin} and theoretical advances in approximate max flow computations have not been yet translated to scalable software implementations \cite{miller2013approximate,peng2016approximate,madry2010fast}, practitioners usually favor a $\frac{1}{2}$-approximation algorithm that uses linear-space and run in linear time for unweighted graphs.  Specifically,  Charikar \cite{charikar2000greedy} showed that the greedy algorithm proposed by Asashiro et al.  \cite{asahiro2000greedily} is a $\frac{1}{2}$-approximation algorithm. We refer to this algorithm as the {\sc Greedy Peeling} algorithm.  
The algorithm removes in each iteration, the node with the smallest degree. This process creates a nested sequence of sets of nodes $V=S_n \supset S_{n-1} \supset S_{n-2} \supset \ldots \supset S_1 \supset \emptyset$. The algorithm outputs the graph $G(S_j)$ that maximizes the degree density $\rho(S_j)$ among  $j=1,\ldots,n$.    Recently, Boob et al. proposed a new algorithm 
\cite{boob2020flowless} that combines near-optimal solutions with time  efficiency.  We emphasize that the DSP has no restrictions on the size of the output. When one imposes cardinality constraints on the size of $|S|$ the problem becomes NP-hard. For instance, solving the DSP with the requirement that the output set should have exactly $k$ nodes is known as the densest-$k$-subgraph problem  and the state-of-the-art approximation algorithm is due to Bhaskara et al. \cite{bhaskara2010detecting}, and provides a $O(n^{1/4+\epsilon})$ approximation in $O(n^{1/\epsilon})$ time. 
Other versions where $|S|\geq k, |S|\leq k$ have also been considered in the literature see \cite{andersen2009finding}.
The DSP has been studied under various other computational models in addition to the RAM model, including dynamic setting \cite{bhattacharya2015space,epasto2015efficient,sawlani2019near}, the streaming model \cite{bahmani2012densest,bhattacharya2015space,mcgregor2015densest,esfandiari2015applications}, and the MapReduce model \cite{bahmani2012densest,bahmani2014efficient}.  In recent years, notable extensions of the DSP have been proposed. Tsourakakis proposed the $k$-clique DSP \cite{tsourakakis2014novel,tsourakakis2015k} that can be solved fast on massive graphs \cite{mitzenmacher2015scalable,sun2020kclist++}, and  Tatti and Gionis \cite{tatti2015density} introduced the \emph{locally-dense} graph decomposition.

 \spara{Tests of fit.} Parametric hypothesis testing  is an extensive area of statistics \cite{wasserman2004hypothesis}.  We are interested in tests of fit for the multinomial distribution, as Benford's law can be thought of as a 9-sided loaded die that determines the first digit of a transaction weight. More generally, we wish to test the hypothesis 
 
 $$ H_0 : F(x) = F_0(x),$$ 
 
 where $F_0(x)$ is some particular cumulative distribution function. This problem is known in general as a {\it goodness-of-fit} problem \cite{smith1981approximating,hoeffding1965asymptotically}. In the case where we have $k$ classes, and we assume $F_0$ corresponds to the multinomial distribution $(p_1,\ldots,p_k)$, if $n_i$ is the number of observations from class $i$ for $i=1,\ldots,k$, and the total number of observations is $n=\sum_{i=1}^k n_i$ then the likelihood function is given by 
 $$ L(n_1,\ldots,n_k\mid p_1,\ldots,p_k) \propto \prod_{i=1}^k p_i^{n_i}.$$
 
 It is well known that the likelihood of alternative hypothesis $H_1: F(x)=F_1(x)$ is maximized when we consider $F_1$ to be the multinomial distribution with parameters given by the ML estimators of its parameters, namely $(\pi_1=\frac{n_1}{n},\ldots, \pi_k=\frac{n_k}{n})$. The likelihood ratio (LR) statistic for testing $H_0$ against $H_1$ is equal to the ratio of the two likelihood functions, and is equal to $\ell = n^n\prod_{i=1}^k \Big(\frac{p_i}{n_i}\Big)^{n_i}$. Hypothesis $H_0$ is rejected when the LR statistic is small enough.   It is known that as $n\to +\infty$, the function $-2\log \ell$ is asymptotically distributed in the $\chi^2$ form with $k-1$ degrees of freedom. Karl Pearson proposed the $\chi^2$ statistic   $X^2 = \sum_{i=1}^k  \frac{(n_i-np_i)^2}{np_i}$. Asymptotically the LR test, and the $X^2$ statistic have the same distribution, but it is known that for small values of $n$ they can yield different results.  
 
 \noindent  \spara{Theoretical Preliminaries.} In Section~\ref{sec:proposed} we use the following  Chernoff bound  \cite{mitzenmacher2017probability}.

\begin{theorem}[Chernoff bound]
\label{th:chernoff} 
Consider a set of mutually independent binary random variables $\{X_1, \ldots, X_t\}$. 
Let $X = \sum_{i=1}^t X_i$ be the sum of these random variables. 
Then, for $0<\epsilon<1$ we have  

$$\Prob{|X-  \Mean{X}| \geq  \epsilon \Mean{X} } \leq 2e^{-\epsilon^2 \Mean{X}/3}.$$ 
\end{theorem}

	\section{AntiBenford Subgraphs}
\label{sec:proposed}

Given a weighted transaction network $G(V=[n],E,w)$, we view the set of its edges $E$ as iid samples from Benford's distribution, and we use the standard $\chi^2$-statistic to test whether it follows such distribution~\cite{hoeffding1965asymptotically}. We refer to this hypothesis as $H_0$.

 The  $\chi^2$-statistic of a subgraph $G[S]$ induced by $S\subseteq V$  is defined as  
 $$\chi^2(S) = \sum_{d=1}^k \frac{ \Big(X_{S,d} -\mathbb{E}(X_{S,d})\Big)^2 }{\mathbb{E}(X_{S,d})},$$ 
 
 \noindent where $X_{S,d}$ is the number of edges in subgraph induced by $S$ whose weight's first digit is $d=1,\ldots,k\myeq 9$, and $\mathbb{E}(X_{S,d})$ is the expected number of  edges whose weight's first digit is $d$ according to Benford's distribution. When it is clear to which subset $S \subseteq V$ we are referring to, we use $X_d$ instead of $X_{S,d}$ for brevity. The  $\chi^2$-statistic of the whole graph $G$ is defined as  $\chi^2(V)$.

 It is a well-known fact this statistic follows the $\chi^2$ distribution with $k-1=8$ degrees of freedom. 
We define the average  $\chi^2$-statistic of the graph as $\psi(V)=\frac{\chi^2(V)}{n}$, and more generally the average $\chi^2$-statistic of any induced subgraph $G[S]$ as $\psi(S) = \frac{\chi^2(S)}{|S|}$, where $\chi^2(S)$ is the $\chi^2$ statistic of the weights induced by $S$.


\begin{definition}[\ab subgraph]
An \ab subgraph is a  subset of nodes $S \subseteq V$  such that $\psi(S) \gg  \frac{e(S)}{|S|}$.
\end{definition}

The above definition is founded on our probabilistic analysis, see Corollary~\ref{cor:ab}. Assuming the  null hypothesis $H_0$ is true, the average $\chi^2$ score of any subgraph is a diminishing fraction of its order:

\begin{align*}
\Mean{ \psi(S) } &= \Mean{ \frac{\chi^2(S)}{|S|} } = \frac{  \Mean{\chi^2(S)} }{|S|} = \frac{1}{|S|} 
 \sum_{d=1}^9 \frac{  \Var{X_{S,d}} }{\Mean{X_{S,d}}} \\
 &= \frac{1}{|S|}  \sum_{d=1}^9 \frac{p_d(1-p_d) e(S)}{p_d e(S)}= \frac{9-\sum_{d=1}^9 p_d}{|S|}=\frac{8}{|S|}.
\end{align*}

However, even if the null hypothesis $H_0$ is true, we do expect to observe subgraphs that deviate from the mean. Roughly speaking, our mathematical analysis shows that the key quantity to look at is the degree density, i.e., half of the average degree. Notice that  a claim of the form ``there does not exist a large subgraph whose average $\chi^2$-score is much larger than its degree density'', is far from obvious; the standard probabilistic approach of bounding the failure probability of a given subgraph by $\frac{1}{n^C}$ for any constant $C$, and a union bound over all possible subgraphs does not yield any meaningful results, since we have an exponential number of bad events. 

Our key theoretical result shows that under hypothesis $H_0$ there exist no \ab subgraphs with average degree $\Omega(\log n)$.

\begin{theorem}
\label{th:ab}
Let $0<\epsilon<1$ be an accuracy parameter, and let $\delta \myeq \min_d p_d=p_9=0.048$ be the lowest digit probability in Benford's law.  Suppose that the null hypothesis $H_0$ is true, and all edge weights are iid samples from Benford's distribution. Then, with high probability for all subgraphs $S \subseteq V$ with average degree at least $C_{\epsilon}\log n$ where $C_{\epsilon}=\frac{36}{\delta \epsilon^2}$ is a constant depending on $\delta, \epsilon$, the number of edges $X_{S,d}$ that start with digit $d$ in $G[S]$ is strongly concentrated around its expectation, for all $d=1,\ldots,9.$
\end{theorem}

\begin{proof}
Consider a subgraph induced by $S\subseteq V$, and a digit $d$. Let $e(S), X_{S,d}$ be the number of induced edges, and the number of edges whose weight starts with digit $d$ respectively. Notice that $\Mean{ X_{S,d} }=p_d e(S)$. By the Chernoff bound (see Theorem~\ref{th:chernoff}) we obtain that for any $0<\epsilon<1$  the following inequality holds

\begin{align*}
\Prob{| X_{S,d}-p_d e(S)|\geq \epsilon p_d e(S)} &\leq 2 e^{-\frac{\epsilon^2 p_de(S)}{3}}\leq 2 e^{-6 |S|\log n} \leq n^{-6|S|}.
\end{align*}

\noindent We used the fact that $\Mean{ X_{S,d}}=p_d e(S)\geq 18|S| \log n$, since the average degree satisfies $\frac{2e(S)}{|S|}\geq \frac{36}{\min_d p_d \epsilon^2}\log n$. Now, consider the following double union bound over all digits, and subgraphs with average degree $\frac{2e(S)}{|S|}=\Omega(\log n)$ as described above: 

\begin{align*}  
\Prob{\exists d \in \{1,\ldots,9\}, S \subseteq V:  X_{S,d} \notin (1\pm \epsilon)\Mean{ X_{S,d}} } &     \\
\leq \sum_{d=1}^9 \sum_{k=2}^n {n \choose k}n^{-6k}  \leq 
9 \sum_{k=2}^n \Big(\frac{en}{k}\Big)^k n^{-6k} =o(1)&\\
\end{align*} 

Therefore, we conclude that with high probability $1-o(1)$, the number of edges with first digit $d$ in subgraphs with large enough average degree is strongly concentrated around the expectation for all nine possible digits $d$.
\end{proof}

This implies that we expect for {\em all} sufficiently large subgraphs to have a strong concentration of $X_{S,d}$ around the true expectation. This fact in turn implies that the average $\chi^2$ score of a set $S$ satisfying the conditions of Theorem~\ref{th:ab} is within a factor of $\epsilon^2/2$ of its average degree. To see why, notice that with high probability the following holds:

\begin{align*}
    \psi(S) &=  \frac{1}{|S|} \sum_{d=1}^9 \frac{(X_{S,d}-p_de(S))^2}{p_de(S)} 
    \leq \frac{1}{|S|} 
    \sum_{d=1}^9 \epsilon^2 p_de(S)=\frac{\epsilon^2}{2} \frac{2e(S)}{|S|}.
\end{align*}

\noindent Since $\epsilon<1$, we state this informally as the next corollary:

\begin{corollary}
\label{cor:ab} 
Under the null hypothesis $H_0$, we expect to observe subgraphs 
$S \subseteq V, |S|=\Omega(\log n)$ with $\psi(S) \approx \frac{e(S)}{|S|}$. Furthermore, $\max_{S \subseteq V}\psi(S) \leq \rho^{\star}$, where $\rho^{\star}$ is the density of the densest subgraph in $G$.
\end{corollary}

\begin{algorithm}[!ht]
	\caption{\label{algo:ab} \ab subgraphs }
	\SetKwInput{Input}{Input}
	\SetKwInput{Output}{Output}
	\Input{\ $G=(V,E, w)$ }  
	\Output{\  Set of nodes $S\subset V$}  
	\For{$u\in V$}{
		$E_u = \{ w(u,v): (u,v) \in E \}$ \;
		$s(v)=\sum_{d=1}^9 \frac{ \Big(X^{u}_d -\mathbb{E}(X_d)\Big)^2 }{\mathbb{E}(X_d)}$
	}
	\For{$(u,v)\in E$}{
		$W_{uv}=\sqrt{ s(v)\cdot s(u)}$ \;
	}
	\tcc{Find the densest subgraph in $G'$}
	$S = DSP(G'=(V,E,W))$\;
	\If{ \textbf{not} $\psi(S) \gg \psi(V)$ } {
		\tcc{No statistically significant anomalous subgraph found}
		Return $\emptyset$  \;
	}
	\Return $S$
\end{algorithm}

\spara{Algorithm.}  Our approach is outlined in pseudocode~\ref{algo:ab}.   For each node $u$, we measure how the transactions it is involved into, deviate from Benford's distribution by computing the $\chi^2$ score. We define the anomaly score $s(u)$ of node $u$ as 

$$ s(u)  = \sum_{d=1}^9 \frac{ \Big(X^{u}_d -\mathbb{E}(X^{u}_d)\Big)^2 }{\mathbb{E}(X^{u}_d)},$$  

\noindent where $X_d^{u}$ is the number of transactions incident to $u$ whose first digit is $d$.   Once we have computed all scores $\{s(u)\}_{u \in V}$ we reweigh each edge by a function of the scores. Specifically, we find that the following function $f(u,v)=\sqrt{s(u)\cdot s(v)}$ works well in practice. Then we solve the densest subgraph problem, namely we maximize the following objective $ \max_{S \subseteq V} \frac{ \sum_{(u,v) \in E(S)} f(u,v)}{|S|}.$ Our choice of reweighing biases the densest anomalous subgraph to contain nodes that are all involved in transactions  that deviate from Benford's; consider a node $u$ with high degree towards an anomalous set of nodes, but assume that overall the transactions that involve $u$ agree with Benford's, so $s(u)$ is low. Then, our reweighing scheme will avoid including such a node in an optimal solution, in contrast to other possible \ab formalizations of the form $\max_{S\subseteq V} \frac{e(S)+\sum_{v\in S} s(v)}{|S|}$. Such formulations are known to be solvable in polynomial time for non-negative scores, see~\cite{goldberg1984finding,hooi2016fraudar}.   Notice that one can use a linear program, a max-flow approach, or a greedy approximation algorithm for finding the DSP (line 6) in the algorithm. Let $\alpha$ and $T_{\text{dsp}}(n,m)=\Omega(n+m)$ be the approximation guarantee ($\alpha=1$ for exact) and the run time of the algorithm DSP used in line 6. Then, we obtain the following straight-forward proposition:

\begin{proposition} 
We can find an $\alpha$-approximate \ab subgraph in $O(n+m+T_{\text{dsp}}(n,m))= O(T_{\text{dsp}}(n,m))$ time in the standard RAM model of computation.
\end{proposition}

 In our experiments, we favor for the scalable greedy algorithm. Finally, in practice we may want to extract more than one \ab subgraph. In order to do so, we run Algorithm~\ref{algo:ab} repeatedly, and each time we remove the previous output set of vertices from the graph, thus finding node-disjoint \ab subgraphs.  
	\section{Experiments}
\label{sec:exp}

\subsection{Experimental Setup}
\label{subsec:setup} 

\spara{Datasets and competitors.} We use Ethereum token transfer networks together with synthetic data to evaluate our proposed framework. As competitors we use EigenSpoke~\cite{Prakash2010eigen}, HoloScope~\cite{liu2017holoscope}, and FlowScope~\cite{li2020flowscope}. 
The real-world datasets are summarized in Table~\ref{tab:datasum}. 


$\bullet$  {\it Ethereum blockchain.} We use the Ethereum token transfer dataset supported and shared by Google BigQuery~\cite{KaggleEthereum}. The original data contains for each token transaction its source and sink addresses, value, timestamp and block information.   We present our anomaly detection results from two snapshots, spanning the periods of January 2018 and January 2019. We filter out all transactions with values less than 1 unit.  We also construct five cumulative Ethereum token transfer networks spanning from one week to the whole year of 2019 to analyze the running time of our method, see the second part of Table~\ref{tab:datasum}.

\begin{table}[h]
\centering
	\caption{Dataset statistics summary}
	\label{tab:datasum}
	\begin{tabular}{ccccc}
		\toprule
		Name & \# nodes & \# edges & \# transactions  \\
		\midrule
		ETH-Jan-18 & 1\,761\,571 & 2\,749\,707 & 4\,279\,799 \\
		ETH-Jan-19 & 2\,199\,347 & 3\,331\,594 & 6\,128\,061 \\
		\midrule
		ETH - 1 Week & 779\,237 & 1\,016\,481 & 1\,438\,641\\
		ETH - 1 Month & 2\,199\,347 & 3\,331\,594 & 6\,128\,061 \\
		ETH - 3 Month & 2\,877\,055 & 8\,504\,999 &  17\,529\,925 \\
		ETH - 6 Month & 5\,018\,047 & 21\,572\,152 &  43\,885\,527 \\
		ETH - 1 Year & 6\,820\,719 & 38\,917\,136 &  85\,055\,054 \\
		\bottomrule
	\end{tabular}
\end{table}

\spara{Machine Specs and code.} The experiments are performed on a single machine, with Intel i7-10850H CPU @ 2.70GHz and 32GB of main memory. Our graph analysis code is written in Python3, and we use the C++ implementation of DSP algorithms from~\cite{boob2020flowless} for efficiency. We use the spartan2\footnote{\url{https://github.com/BGT-M/spartan2}} library for all competitors. The code and datasets are available at \url{https://github.com/tsourakakis-lab/antibenford-subgraphs}.

\subsection{Synthetic data}
\label{subsec:edgeweights} 

The goal of our experiments on synthetic data is two-fold. First, we want to show that   popular competitors cannot detect the planted anomalies, and secondly show the limitations of \ab subgraphs. 

Towards this purpose, we generate clustered networks, with planted anomalies with respect to the first digit distribution in order to compare how well various methods can detect them.  Since the original HoloScope implementation assumes that the input has a bipartite structure, we generate a version of the stochastic blockmodel as follows. We have two types of nodes (users vs. objects), and nine clusters. Each cluster is a bipartite clique, and clusters are interconnected with edge probability $p=0.1$, namely for each pair of user $u$, and object $i$ from two separate clusters we generate an edge independently with probability $p$. Then we sample first digits on the edges (since we only care about the first digit, this suffices) according to Benford's distribution for all the edges {\it except for} the  induced edges for three clusters. Those three bicliques receive digits 1, 2, 3 with probability 1 respectively. 
We set the size of each normal cluster to 80 nodes, and  we range the size of the anomalous bicliques in [20, 50, 80, 110].  
 
\begin{figure}[h]
    \centering
    \begin{tabular}{c}
    \includegraphics[width=0.8\textwidth]{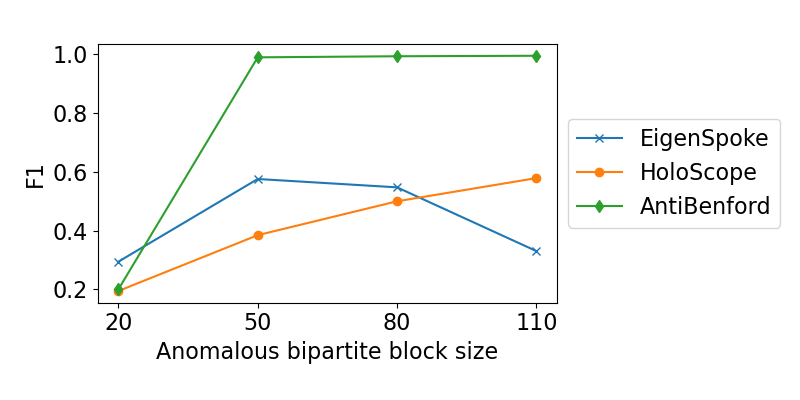}
    \end{tabular}
    \caption{\label{fig:synth} F1 score  for the synthetic Antibenford experiments. Our method  --in contrast to the competitors HoloScope and EigenSpoke--   can detect the anomalous clusters, when their size is greater than 20. For details, see section~\ref{subsec:edgeweights}.}
\end{figure}

Figure~\ref{fig:synth} shows the results of the three methods. Eigenspoke and Holoscope cannot detect the planted anomalous subgraphs. For example, Holoscope outputs the whole graph. It is worth emphasizing that the increase in the F1 of Holoscope is due to the fact that the size of the anomalous clusters increases, and not due to any changes in the performance of the algorithm. Our method achieves perfect precision and recall when the size of the anomalous clusters is greater than 20, by finding the top-3 outputs of Algorithm~\ref{algo:ab}. This is done by running the algorithm, removing the output subgraph from the graph, and iterating for three times. This shows a well-known limitation of the densest subgraph problem, pointed out by prior work~\cite{tsourakakis2013denser,tsourakakis2014novel,tsourakakis2015k}, namely the small size of the true anomalous clusters is ``hidden'' by larger sets of higher average degree. This issue can be resolved by using a weighted version of the $(\alpha,\beta)$-biclique DSP problem \cite{mitzenmacher2015scalable,tsourakakis2015k} with $\alpha, \beta >1$.

\subsection{Anomaly Detection in Ethereum Networks}
\label{subsec:eth} 

\begin{figure*}[h]
	\centering
		\includegraphics[width=.32\linewidth]{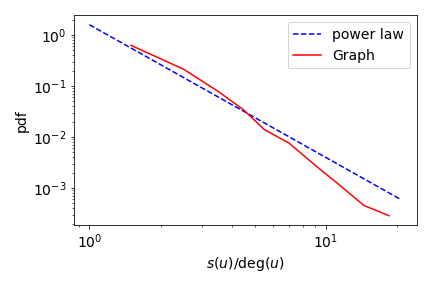}
		\includegraphics[width=.32\linewidth]{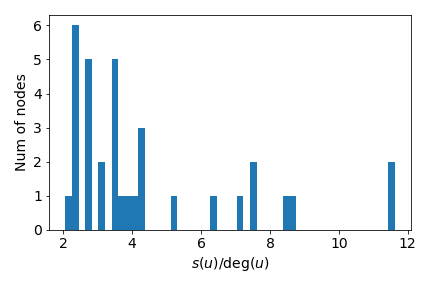}
		\includegraphics[width=.32\linewidth]{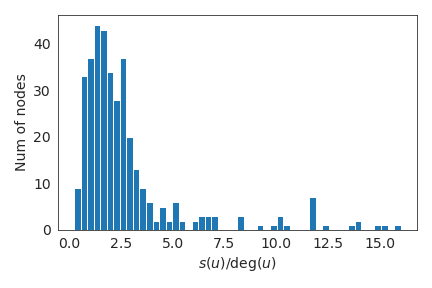}
		\caption{Left: Histogram of the quantity $s(u)\over \deg(u)$ of nodes in (a) ETH-Jan-18, which roughly fits power law with $\alpha=2.6$. Middle, right: Histogram of $s(u)\over \deg(u)$ of nodes in two \ab subgraphs.}
		\label{fig:powerlaw}
\end{figure*}

 In this section we analyze two Ethereum transaction networks, ETH-Jan-18 and ETH-Jan19, and demonstrate the suspicious subgraphs we find. By iteratively finding and removing suspicious subgraphs by applying Algorithm~\ref{algo:ab}, we find the top-5 \ab subgraphs and show their $\chi^2$ and $\psi$ scores in Table~\ref{tab:chi_stat_eth}, together with the statistics for the whole network that closely follows Benford's law. We observe a sheer contrast of the value of $\psi$  between the global network and the suspicious subgraphs we found. In Table~\ref{tab:chi_stat_eth} we also mark highly suspicious subgraphs whose $\psi$ values are significantly higher than their densities $|E|/|S|$, according to our rigorous analysis, see Corollary~\ref{cor:ab}. Figure~\ref{fig:powerlaw} shows the empirical distribution of the quantity $s(u)/deg(u)$, i.e. the $\chi^2$ score of node $u$ averaged by its degree. The histogram of this quantity is skewed, and follows (roughly) a power law with slope $\alpha = 2.6$ on the whole network of ETH-Jan-18~\cite{alstott2014powerlaw}. We obtain this fitting using the Python package {\it powerlaw} \cite{alstott2014powerlaw}. The middle and right figures in Figure~\ref{fig:powerlaw} show the histograms for two top \ab subgraphs that deviate from the histogram of the global graph on the left.

\begin{table}[]
\small
    \caption{Total and average $\chi^2$ statistics of top five \ab subgraphs found on the ETH networks. Subgraphs whose $\psi$ is greater than $e(S)/|S|$ are marked yellow and highly suspicious according to Corollary~\ref{cor:ab}.}
    \label{tab:chi_stat_eth}
    \centering
    \begin{tabular}{c|c|ccccc|c}
        Dataset & Metrics & $1^{st}$ & $2^{nd}$ & $3^{rd}$ &  $4^{th}$ & $5^{th}$ & global\\
        \midrule
        ETH- & $\chi^2$   & 3429 & 2587 & 379 & 2.59e5 & 698 & 1.6e5 \\
        Jan-18 & $\psi$  & 14.4 & 7.59 & \cellcolor{yellow}11.14 & \cellcolor{yellow}35.75 & \cellcolor{yellow}5.96 & 0.09 \\
        & $|E|/|S|$ & 18.8 & 13.09 & 3.82 & 1.96 & 1.99 & 2.43 \\
        \midrule
        ETH- & $\chi^2$   & 131818 & 3963 & 1205 & 10677 & 9372 & 2.92e5\\
        Jan-19 & $\psi$  & \cellcolor{yellow}1588 & 10.88 & \cellcolor{yellow}9.06 & 9.58 & \cellcolor{yellow}9.49 & 4.26e-3 \\
        & $|E|/|S|$ & 214 & 15.5 & 4.77 & 14.15 & 3.96 & 2.79 \\
    \end{tabular}
\end{table}

\begin{figure}[h]
	\centering
		\includegraphics[width=.5\linewidth]{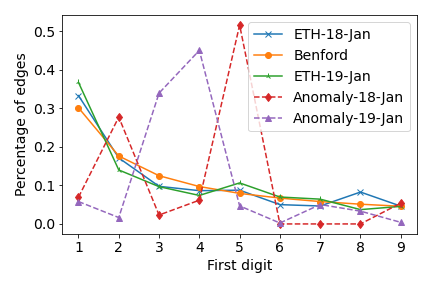}
		\caption{First digit distribution of Benford's law,  transaction values of ETH-Jan-18 and ETH-Jan-19 networks, and the suspicious subgraphs found.}
		\label{fig:eth2018_19_dist}
\end{figure}

Figure~\ref{fig:eth2018_19_dist} describes the first digit distribution of the subgraph we found from the Ethereum token transfer network in January 2018. Transaction values of this subgraph strongly violate Benford's law as half of them start with 5. Figure~\ref{fig:enron_intro}(b) visualizes the topology which is close to a tripartite directed graph. According to previous studies \cite{starnini2021smurf,threestage,li2020flowscope}, this follows a common money laundering network pattern where money flows {\em quickly} (i.e., in a matter of a couple of weeks) from few sources to few targets via middle nodes. We can   even visually recognize the sources, middle nodes, and final recipients. Over $140\,000$ Ethereums are transferred from top to bottom in one month in this subgraph. Interestingly we verify that all middle nodes send out all the money they receive, which is a property known to hold in money laundering as the  \textit{zero out} property, see also~\cite{li2020flowscope,starnini2021smurf}.  A sample of such transactions are shown in Table~\ref{tab:eth_2018trans}. We observe that a certain amount of money is sent to several middle accounts on the same day by one sender, and is altogether sent out to one receiver after at most a couple of weeks, and sometimes even earlier than that.

\begin{table}[h]
	\centering
	\begin{tabular}{c|c|c|c|c}
		\centering
		\begin{tabular}{@{}c@{}}Sender \\ address\end{tabular}   & \begin{tabular}{@{}c@{}}Receiver \\ address\end{tabular} & \begin{tabular}{@{}c@{}}Amount (ETH)  \\ per middle \\ account\end{tabular} & \begin{tabular}{@{}c@{}}In Date \\ (yy/mm/dd)\end{tabular}  & \begin{tabular}{@{}c@{}}Out Date \\ (yy/mm/dd)\end{tabular} \\
		\hline
		beb58d36 & 9b06578f & 9667                   & 18/01/18    & 18/02/06     \\
		07c15792 & 59c42026 & 2160                   & 17/09/15    & 17/09/24     \\
		9a974c3b & 59c42026 & 33565                  & 17/10/13    & 17/10/20     \\
		c0cb31d0 & 89705733 & 17200                 & 18/02/03    & 18/02/17    
	\end{tabular}
	\caption{\label{tab:eth_2018trans}Suspicious transactions started from one sender, through multiple middle accounts, and ended at a receiver at the same date, graph structure visualized in Figure~\ref{fig:enron_intro}(a). We only show the last eight characters of address hash. In and out date correspond to the day Ethereums are transferred into and out of middle accounts respectively.}
\end{table}

\begin{figure}[h]
	\centering
	\begin{tabular}{cc}
				\includegraphics[width=0.5\linewidth]{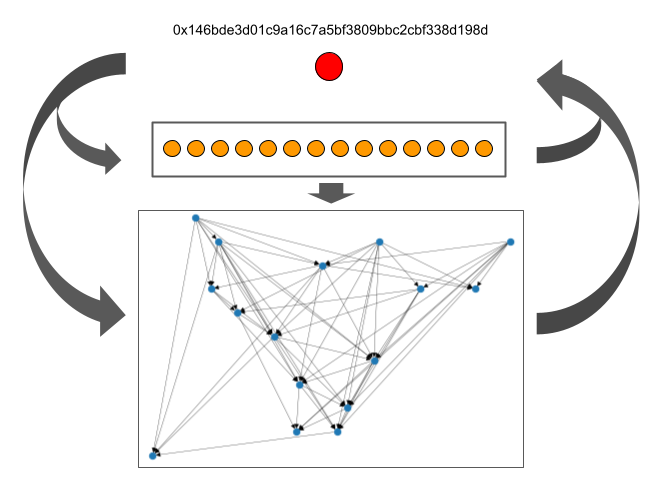} & \includegraphics[width=0.5\linewidth]{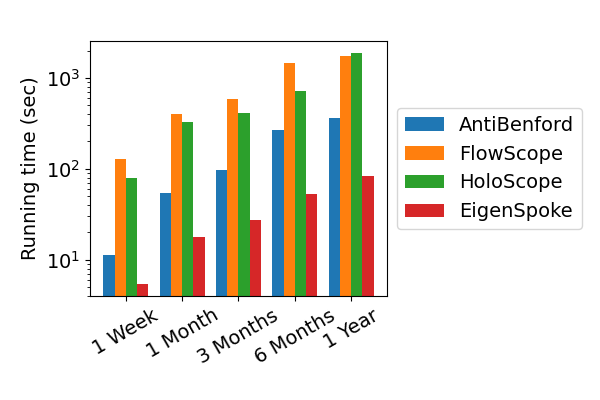}\\
				(a) & (b) \\ 
\end{tabular}
		\caption{		\label{fig:eth2019_struc} (a) Suspicious subgraph found from Ethereum token transfer network of Jan 2019. \textit{Center} node circulates more than $0.7$ million USD from Oct 2018 to Mar 2019. (b) Running time comparison of AntiBenford subgraphs and other methods on cumulative snapshots of Ethereum transaction networks.}
\end{figure}

On the Ethereum Jan. '19 network, our method outputs a clearly anomalous subgraph. The distribution of its digits significantly deviates from Benford's law, which, as we mentioned before, fits well the empirical global first digit distribution. Most of the transactions start with digits 3, and 4. Table \ref{tab:chi_stat_eth} summarizes the statistics of the top-5 anomalous subgraphs found using Algorithm~\ref{algo:ab}, and Figure~\ref{fig:eth2018_19_dist} shows the resulting digit distributions, contrasted with the global digit distributions that follow closely Benford's law. Figure~\ref{fig:eth2019_struc}(a) shows that the subgraph can be partitioned into three groups of nodes. There is a central node, depicted as red, a set of middle nodes marked as orange, and a group of accounts marked blue that form a more complicated network, but still exhibit a DAG-like structure, with top-down edge flow. Each big arrow between groups represents complete edge connections between two groups in a certain direction. The \textit{central} node in one month sends out more than $1.1$ million Ethereum Tokens to other accounts inside this subgraph, and receives more than $1.2$ Ethereum Tokens from them with some external incomes through the blue bottom accounts.

Flowscope -a framework tailored for anti-money laundering- is unable to output any of these subgraphs. Even worse, its outputs do not appear suspicious at all, and consist of a path of length 2, with large weights on the edges. Furthermore, we observe that suspicious structures  in real-world financial networks do not always exhibit a perfect $k$-partite structure, and thus are not easily detected by Flowscope. 

 \spara{Scalability.} We apply \ab and other competitors on a set of five Ethereum networks that span transactions from one week to one year. We report the running time in Figure~\ref{fig:eth2019_struc}(b). Due to the linear complexity of greedy peeling, Algorithm~\ref{algo:ab} can efficiently find anomalies even in the largest graph we have.

	\section{Conclusion}
\label{sec:conclusion}

In this work we contribute a novel, statistically founded formulation for finding anomalies in transaction networks; it is known that unsupervised anomaly detection in financial networks is a challenging, and important problem world-wide. 
We show that \ab subgraphs exhibit characteristics that are known in the literature to appear in illicit transactions, such as smurf-like structures showing that criminals are likely to operate abnormally both in terms of creating special subgraph structures, and splitting the money carefully. We provide  an extension  of Benford's law on networks, and we show that the analysis of such a law presents technical challenges due to the fact that there exists an exponential number of ``bad events''. Based on our analysis we provide a well-founded definition of \ab subgraphs.  We show empirically that the resulting \ab subgraphs are able to find anomalies in Ethereum networks that are otherwise unobtainable from state-of-the-art graph-based anomaly detection methods.  An interesting direction is the design of algorithms that allow for overlapping anomalous subgraphs, and extending the experimental setup to include other measures of statistical deviation.

	\bibliographystyle{abbrv}
	\bibliography{ref}

\end{document}